\pdfoutput=1
\documentclass[fleqn,usenatbib,usedcolumn]{mnras}
\usepackage[british]{babel}             
\usepackage{txfonts}                    
\usepackage[T1]{fontenc}                
\usepackage{graphicx}                   
\usepackage{color}
\usepackage{amsmath}

\hypersetup{pdfauthor={D. Izquierdo-Villalba},
               pdftitle={Radio galaxies},
               pdfkeywords={galaxies:high-redshift -- galaxies:evolution -- methods:numerical},
               bookmarksnumbered=true}
\volume{{\rm submitted}}


\newcommand{\cm}{$\mathrm{C_{M_{S}}\,}$}
\newcommand{\ch}{$\mathrm{C_{M_{H}}\,}$}

\newcommand{\RGo}{$\rm{RG}^{AGNoff}$} 
\newcommand{\cmo}{$\mathrm{C_{M_{S}}^{AGNoff}\,}$ }
\newcommand{\cho}{$\mathrm{C_{M_{H}}^{AGNoff}\,}$ }
\title[Radio galaxies environment]{The environment of radio galaxies: A signature of AGN feedback at high redshifts}

\author[Izquierdo-Villalba D. et al.]{
David Izquierdo-Villalba$^{1}$\thanks{E-mail: dizquierdo@cefca.es},
\'{A}lvaro A. Orsi$^{1}$,
Silvia Bonoli$^{1}$, 
Cedric G. Lacey$^{2}$, \newauthor \space Carlton M. Baugh$^{2}$, 
Andrew J. Griffin$^{2}$
\\
$^{1}$ Centro de Estudios de F\'{\i}sica del Cosmos de Arag\'{o}n (CEFCA), Plaza San Juan 1, Planta-2, Teruel, 44001, Spain.\\
$^{2}$ Institute for Computational Cosmology, Department of Physics, University of Durham, South Road, Durham, DH1 3LE, UK.
}

\begin{document}
\maketitle

\begin{abstract}
We use the semi-analytical model of galaxy formation \texttt{GALFORM} to characterise an indirect signature of AGN feedback in the environment of radio galaxies at high redshifts. The predicted environment of radio galaxies is denser than that of radio-quiet galaxies with the same stellar mass. This is consistent with observational results from the CARLA survey. 
Our model shows that the differences in environment are due to radio galaxies being hosted by dark matter haloes that are $\sim 1.5$ dex more massive 
than those hosting radio-quiet galaxies with the same stellar mass. By running a control-simulation in which AGN feedback is switched-off, we identify AGN feedback as the primary mechanism affecting the build-up of the stellar component of radio galaxies, thus explaining the different environment in radio galaxies and their radio-quiet counterparts. The difference in host halo mass between radio loud and radio quiet galaxies translates into different galaxies populating each environment. We predict a higher fraction of passive galaxies around radio loud galaxies compared to their radio-quiet counterparts. Furthermore, such a high fraction of passive galaxies shapes the predicted infrared luminosity function in the environment of radio galaxies in a way that is consistent with observational findings. Our results suggest that the impact of AGN feedback at high redshifts and environmental mechanisms affecting galaxies in high halo masses can be revealed by studying the environment of radio galaxies, thus providing new constraints on galaxy formation physics at high redshifts.

\end{abstract}


\section{INTRODUCTION}
Within the current picture of galaxy formation, an active galactic nucleus (AGN) is associated with the energy release resulting from  gas accretion onto supermassive black holes (SMBH) 
residing at the center of most galaxies \citep{Soltan1982,Kormendy1995,Richstone1998,Kormendy2001}. Nuclear activity is thought to impact star formation through different
physical processes (such as the heating and compression of the IGM), to which the community using with the general term ``AGN
feedback'' \citep{Silk1998,Birzan2004,Springel2005,DiMatteo2005,Diamond-StanicandRieke2012,Mullaney2012a,Mullaney2012b,Chamani2017,Eisenreich2017,Shabala2017}.

In galaxy formation models, AGN feedback is invoked to heat the gas content of massive galaxies and their host dark matter haloes, thereby quenching star formation and regulating the abundance  of bright, 
massive galaxies \citep[see, e.g.,][]{Benson2003,Granato2004,Bower2006,Croton2006,Cattaneo2008}. In this context, theoretical models typically distinguish between two types of 
AGN feedback: the radiative ``quasar mode'', associated with episodes of efficient cold gas accretion onto the central BH, which is typically triggered by galaxy mergers or disc 
instabilities, and the ``radio mode'', which depends directly on gas accretion from the hot halo surrounding galaxies and is responsible for powering relativistic 
jets \citep{Bower2006,Croton2006,Somerville2008,Cattaneo2008,Lagos2008,Fanidakis2012,Fanidakisnew2013,Henriques2015}. Recent hydrodynamical simulations have shown that AGN feedback can shape 
the central mass distribution of their host galaxies and induce the quenching of star formation \citep{DiMatteo2005,Booth2009,Bonoli2016,Dubois2016,Spinoso2017,Weinberger2017}.

Despite its key role in models, AGN feedback and its impact on galaxy evolution is not well characterised observationally. Studies have not reached conclusive results when they tried 
to explore correlations between AGN luminosity (in the X-ray, optical or radio bands) and host properties such as star formation rates (SFR) or BH accretion 
\citep{Hardcastle2006,Best2007,Shao2010,Georgakakis2011,Harrison2012,Mullaney2012,Rodighiero2015,Stanley2015,Lanzuisi2017,Soergel2017}. Furthermore, the study of feedback processes like 
outflows and winds is extremely difficult to observe \citep[see, e.g.,][]{ Bischetti2017}.\\

Some recent studies have focused on high-$z$ radio-loud AGN (RLAGNs) to prove the existence of a coevolution between the AGN and their host galaxies \citep{Holt2008,Nesvadba2008}. RLAGNs 
are expected to be good candidates to trace AGN feedback since they i) sample some of the most massive galaxies at high-$z$ \citep{Seymour2007}; ii) lie in overdense regions 
\citep{Hill1991,Pascarelle1996,Best2000,Kurk2000,Kurk12004,Kurk22004,Venemans2004,Hatch2011,Hatch2014,Cooke2014,Orsi2015};
and iii) are associated with energetic outflows of ionized gas powered by their central SMBHs \citep{Nesvadba2008,Nesvadba2017}. Recently, the CARLA survey has targeted the 
environment of high-$z$ RLAGNs \citep{Wylezalek2013,Wylezalek2014, Cook2015, Cook2016}. Their results show that radio galaxies (RGs), a sub-sample of RLAGNs, lie in denser environments with respect to radio-quiet galaxies with the same stellar mass \citep{Hatch2014}. This result suggests a link between the environment and  the physical processes connected with the radio 
activity. \\

Here, we explore this link from a theoretical perspective. We use the \texttt{GALFORM} galaxy formation and evolution model \citep{Cole2000,Bower2006,Lacey2016} to study the environment of radio loud and 
radio quiet galaxies and explore the physical mechanisms that lead to differences in the environment of the two  populations. We compare the model predictions with the results obtained from 
the CARLA survey. As the model includes AGN feedback, we can explore how black hole growth affects the  evolution of massive galaxies at high redshifts.\\

The outline of this work is as follows. In \hyperref[sec:GALFORM]{Section~\ref{sec:GALFORM}}, we briefly describe the galaxy formation model used. In \hyperref[sec:Data]{Section~\ref{sec:Data}}, we
study the predicted environment of radio active and radio quiet galaxies and investigate how the differences are due to the role of AGN feedback. In
\hyperref[sec:Quenching around radio galaxies]{Section~\ref{sec:Quenching around radio galaxies}}, we focus on the properties of galaxies surrounding RGs. Finally, in 
\hyperref[sec:Summary]{Section~\ref{sec:Summary}} we summarise our
main findings. Magnitudes are given in the AB system and distances in comoving units. 
\begin{figure*}
	\centering
	\includegraphics[width=0.498\textwidth]{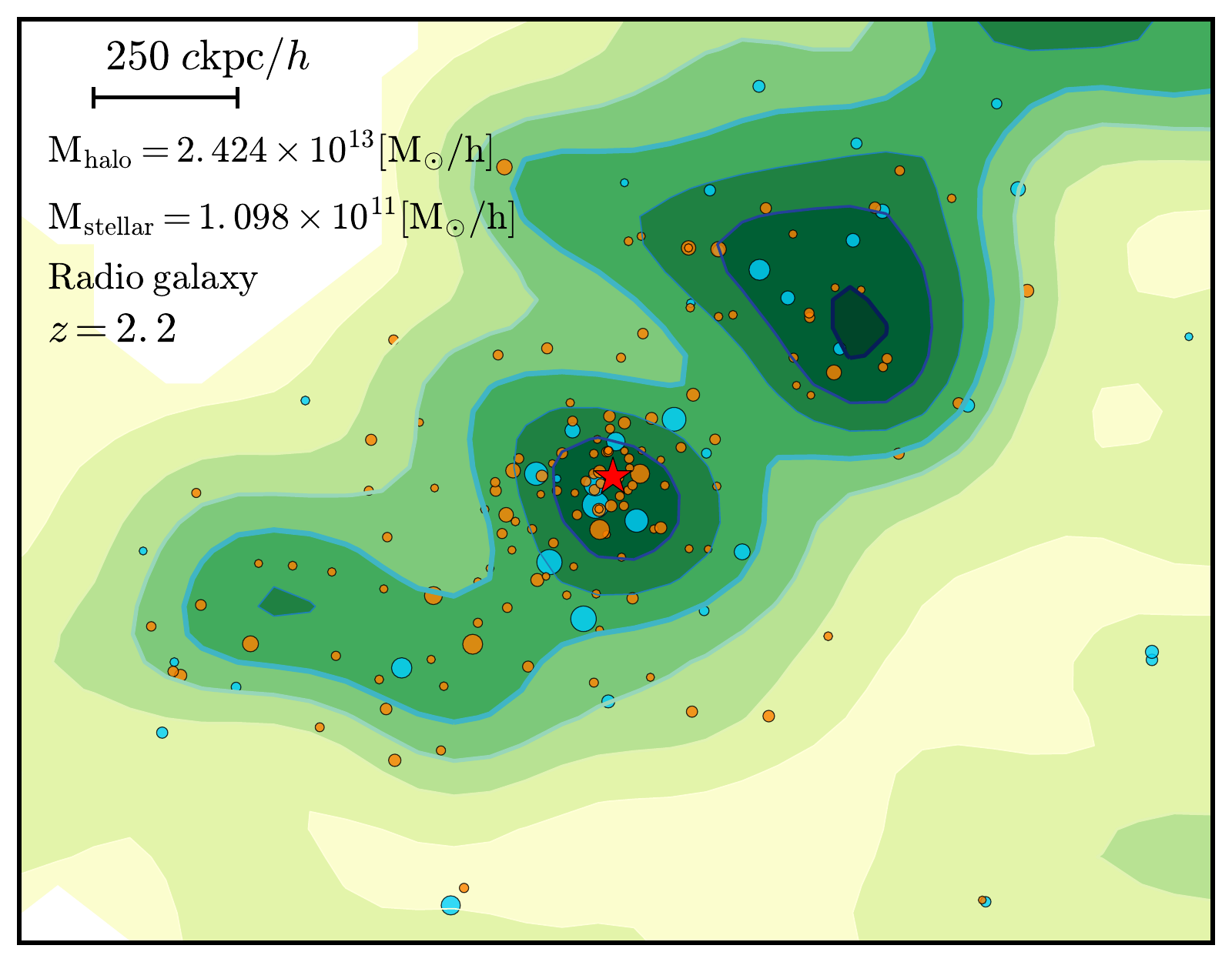} \includegraphics[width=0.498\textwidth]{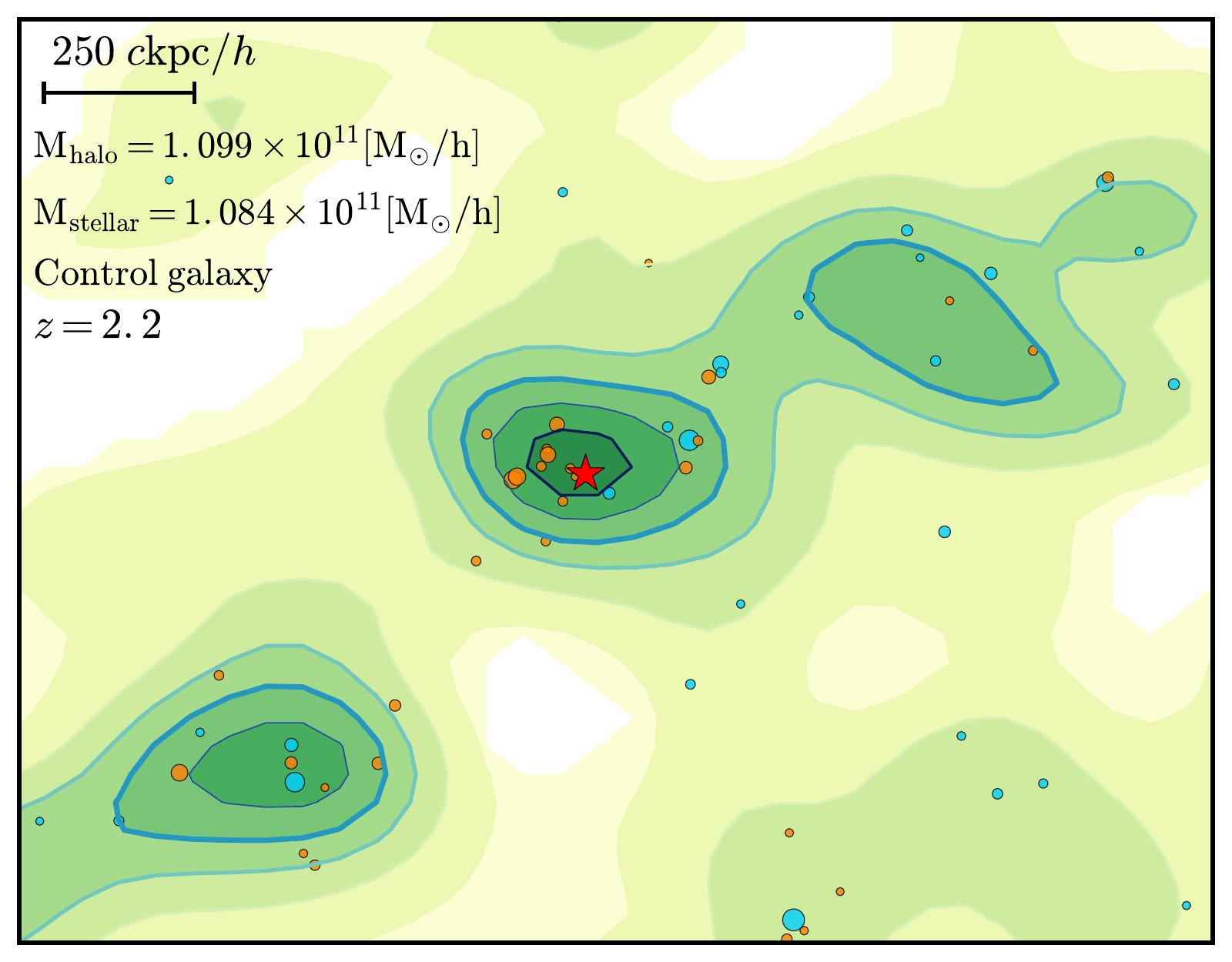}
	\includegraphics[width=0.498\textwidth]{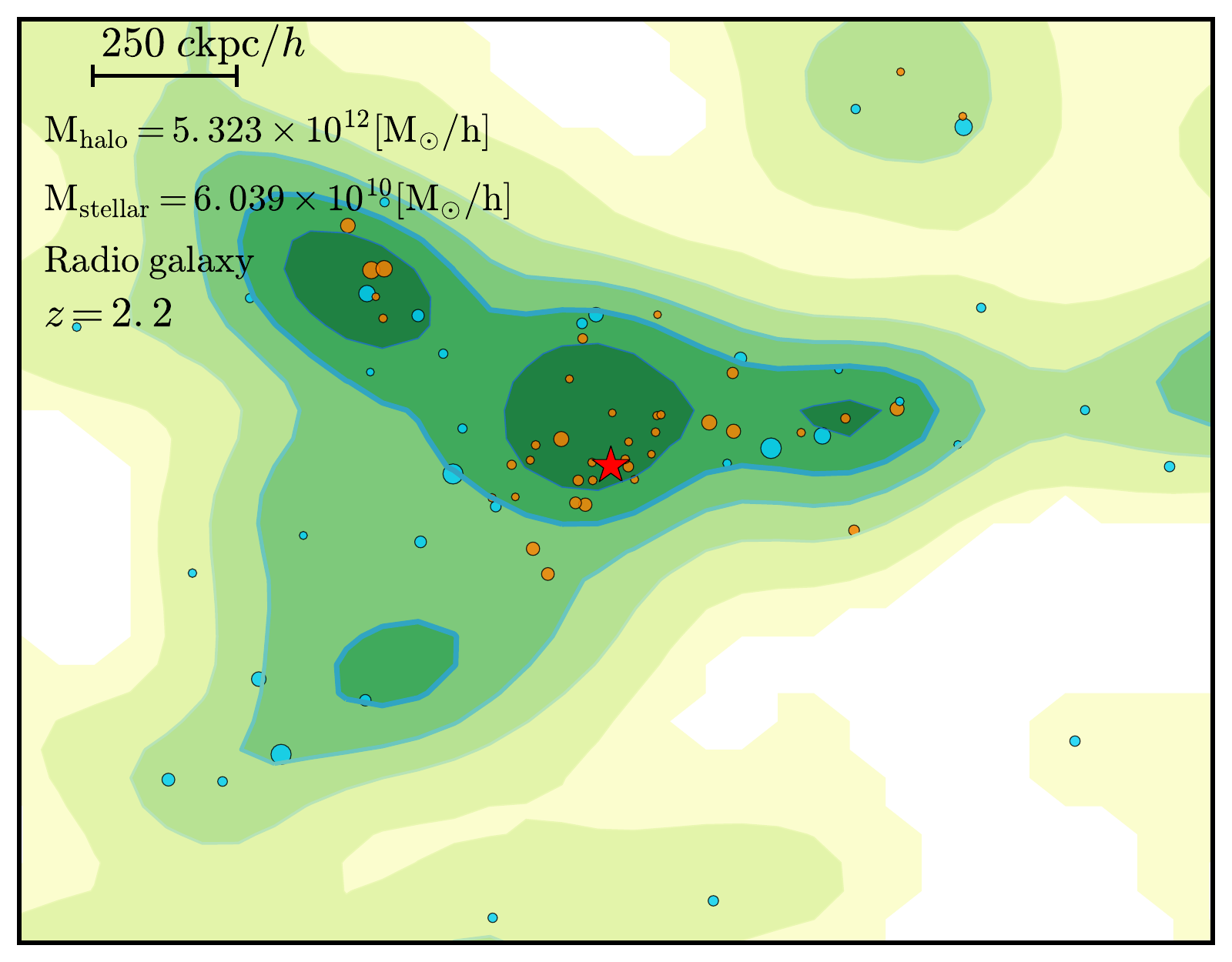} \includegraphics[width=0.498\textwidth]{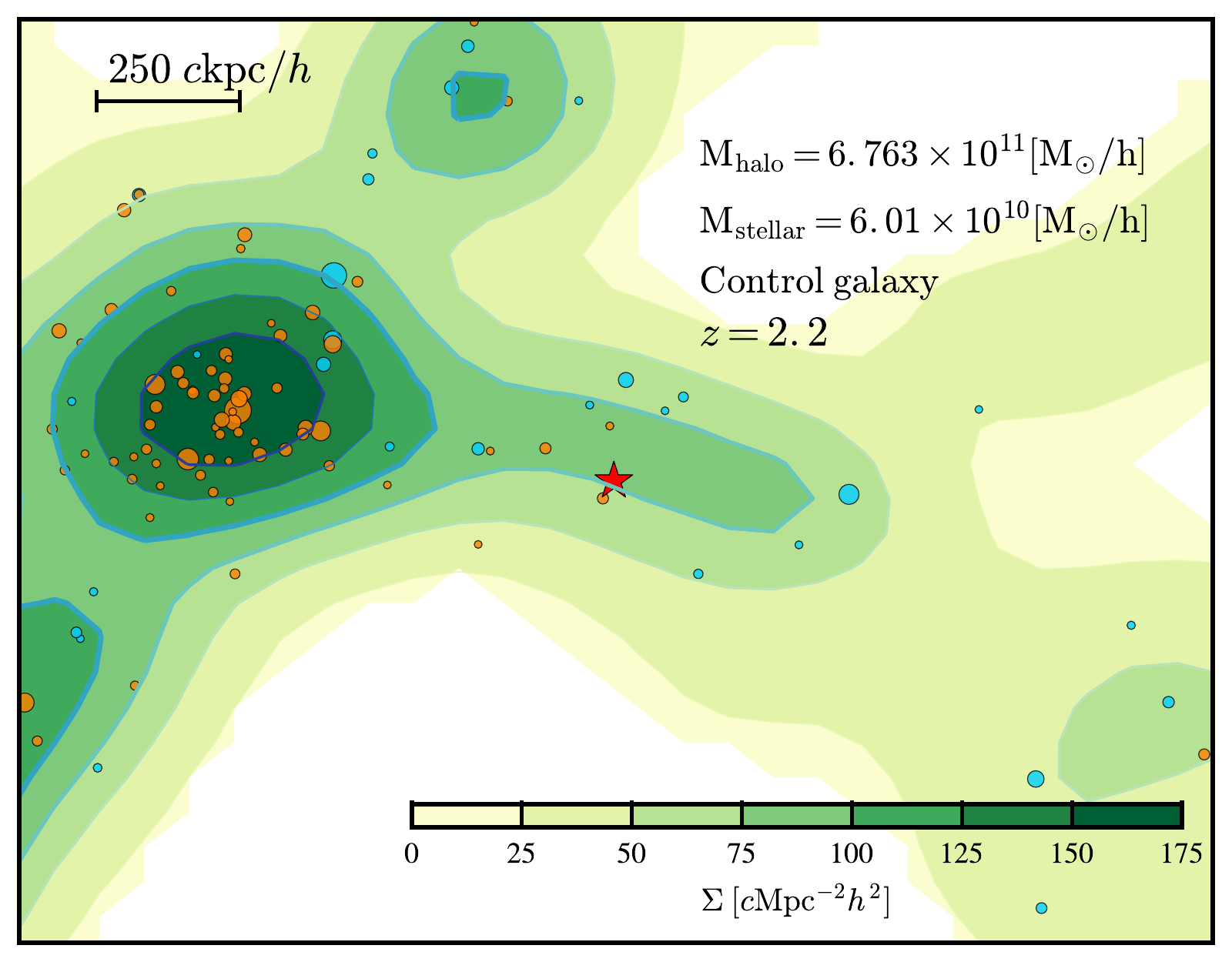}
	\caption{Examples of the environment of two RGs (left
		panels) and two galaxies from the control \cm sample (right panels) at $z$ = 2.2. The
		red stars located at the centre of each panel indicate the position of the RG or
		the \cm galaxy, with stellar and host halo masses  indicated on the panels. Circles indicate the positions of
		neighbouring galaxies, with the symbol  size being proportional to the stellar mass of
		each galaxy, while the colour encodes the specific star formation rate (sSFR) of
		the galaxy: orange circles indicate passive galaxies, while cyan symbols indicate  star-forming
		galaxies, where the threshold between the two populations has been set to a sSFR
		value of $10^{-10} yr^{-1}$. The contours represent the projected surface density of subhaloes around the central galaxy. The thickness of the projection is 5 $c$kpc/$h$ and the length scale is indicated by the legend.}
	\label{fig:Stellardist}
\end{figure*}

\section{GALAXY FORMATION MODEL} \label{sec:GALFORM}
Throughout this work we make use of the semi-analytical model of galaxy
formation \texttt{GALFORM}. A full description of the model can be found in
\citet{Cole2000,Benson2003,Baugh2005,Bower2006} and \citet{Lacey2016}. We use
here the latest variant of the model, described in \citet{Lacey2016} and the
modifications presented in \cite{Baugh2018} and \cite{Griffin2018}. In
brief, \texttt{GALFORM} computes the formation and evolution of the galaxy
population set in a hierarchical structure formation scenario. The main
physical processes driving galaxy formation and BH evolution include: gas
cooling and disk formation in dark matter (DM) halos, leading to star formation
in the disk component and to subsequent regulating mechanisms such as supernova
and AGN feedback; chemical enrichment of gas and stars; galaxy mergers and disk
instabilities leading to bursts of star formation and the formation of a
spheroid component; the evolution of SMBHs and the computation of observed
properties such as AGN and galaxy luminosities. The \texttt{GALFORM} variant in
\citet{Lacey2016}  shows good agreement with a wide range of galaxy proprieties, 
such as their luminosity and stellar mass function, the evolution of Lyman break
galaxies,  galaxy sizes, and the number counts of submillimetre galaxies 
at $z > 2$, among others.

In this work, we run \texttt{GALFORM} in the P-Millennium N-body
simulation \citep{Baugh2018}. This large simulation has a halo mass resolution  of
$\mathrm{2.12 x 10^{9}\;[M_{\odot}/}h]$, corresponding to 20 simulation particles, each with
mass $\mathrm{1.06 x 10^{8}\;[M_{\odot}/}h]$, a periodic box of
$\mathrm{542.16\;[Mpc/}h\rm]$ and cosmological parameters consistent with the
latest cosmological constraints from the {\it Planck} mission
\citep{PlanckCollaboration2016}: $\mathrm{\Omega_M = 0.307}$,  $\mathrm{\Omega_b
= 0.0483}$, $\mathrm{\Omega_\Lambda = 0.677}$, $\mathrm{n_s = 0.968}$,
$\mathrm{\sigma_8 = 0.8288}$ and $h \rm = 0.677$. The model parameters are
re-tuned slightly compared with those used in \citet{Lacey2016}, to compensate
for the change in the cosmological parameters, the improved simulation
resolution and a new galaxy merger scheme \citep{Campbell2015,Simha2016,Baugh2018}.\\

The modelling of SMBH in \texttt{GALFORM} was introduced in
\cite{Malbon2007} and then extended to include AGN feedback by
\cite{Bower2006}, \cite{Fanidakis2012,Fanidakisnew2013} and \cite{Griffin2018}. The model includes three channels in which BHs can grow:
\textit{mergers} with other BHs, and via gas accretion during the
\textit{starburst mode} (or \textit{quasar mode}) and the \textit{hot-halo mode} (or \textit{radio mode}). The \textit{starburst} mode is triggered by disk instabilities or galaxy mergers; during these processes, a large amount of cold gas is expected to be driven
towards the inner parts of the galaxy providing fuel for the BH. The \textit{hot-halo}
mode assumes that the gas is accreted onto the BH directly from the diffuse
quasi-static hot gas atmosphere of the DM halo, without being cooled onto the
galactic disk. To  prevent the formation of too many massive galaxies,
\texttt{GALFORM} invokes the \textit{hot-halo} accretion and the associated AGN feedback as the source of heat which halts gas cooling. There are two conditions under which the AGN feedback starts to quench efficiently the star formation in a galaxy: i) the cooling time of hot gas has to be larger than the free-fall time and, ii) the BH accretion needs to be very sub-Eddington ($\rm f_{Edd}<0.01$) to balance the radiative luminosity of the cooling flow \citep{Bower2006,Fanidakis2010,Lacey2016}.

The code computes the gas accreted by the BH via the \textit{starburst} and \textit{hot-halo} modes (if the AGN feedback conditions are satisfied) at every timestep, and converts it into an accretion rate $\rm \dot{M}$. In the \textit{starburst} mode, the duration of the accretion episode is proportional to the dynamical time-scale of the host spheroid while in the \textit{hot-halo} mode it is computed using the timestep over which the gas is accreted from the halo atmosphere. 
The accretion disc bolometric luminosity ($\rm L_{bol}$) is calculated with the Shakura-Sunyaev thin disk (TD) solution \citep{ShakuraSunyaev1973} if the value of $\rm \dot{M}$ in Eddington units ($\dot{m} \rm = \dot{M}/\dot{M}_{Edd}$) exceeds a critical accretion rate of $\dot{m}_c \rm = 0.01$.

On the other hand, when $\dot{m}\leq \dot{m_c}$, then the advection dominated
accretion flow (ADAF) thick disc solution is adopted \citep{Narayan1994}. The
thin disc accretion channel is linked with the fast rate of BH growth while the
ADAF is linked with slow rate of growth and usually connected with the feedback phase implemented in the model. 
Finally, if the accretion rate becomes supper-Eddington, the bolometric
luminosity is limited to a  factor proportional to Eddington luminosity \citep[see][]{Fanidakis2010, Griffin2018}.

The accretion flow forms a disc around the BH which is able to produce a relativistic jet whose power depends strongly on the disc structure (TD or ADAF), the BH mass and its spin \citep{Meier2002,Fanidakis2010}: 

\begin{align}
\label{equation:jet}
L_{\rm jet,ADAF} & = & 2\times10^{45}\left( \frac{\rm M_{BH}}{\rm 10^9 M_{\odot}}\right) \left( \frac{\dot{m}}{\dot{m_c}}\right) a^2 \;\;\;\;\;\; [{\rm erg}/s], \\ \label{equation:jet2}
L_{\rm jet,TD} & = &  2.5\times10^{43}\left( \frac{M_{BH}}{10^9 M_{\odot}}\right)^{1.1} \left( \frac{\dot{m}}{\dot{m_c}}\right)^{1.2} a^2 \;\;\; [{\rm erg}/s], 
\end{align} where $M_{BH}$ and $a$ are, respectively, the BH mass and spin. In the
super-Eddington regime, it is  assumed that the flow remains in a thin disc state as there is as yet no model to describe the behaviour of the radio jet in this regime.

The jet luminosities can be related to radio luminosities ($\mathrm{L_{\nu R}}$)
using the non-linear dependence between the jet power and black hole mass
($\mathrm{M_{BH}}$) and accretion rate ($\dot{m}$) parameters \citep{Heinz&Sunyaev2003}: 
\begin{align} \label{equation:Radiojet}
\nu_R L_{\nu R,\rm ADAF} & = & A_{\rm ADAF} L_{\rm jet,ADAF}\left(\frac{M_{BH}}{10^9 M_{\odot}}\frac{\dot{m}}{\dot{m_c}}\right) ^{0.42} \\ \label{equation:Radiojet2}
\nu_R L_{\nu R,\rm TD} & = & A_{\rm TD}     L_{\rm jet,TD}  \left(\frac{M_{BH}}{10^9 M_{\odot}}\right)^{0.32}\;\left( \frac{\dot{m}}{\dot{m_c}}\right)^{-1.2} \;, 
\end{align}
where the normalisation factors $A_{TD}$ and $ A_{ADF}$ are adjustable parameters, set to 0.8 and $2\times 10^{-5}$ respectively, in order to match the radio luminosity function at $z = 0$.

The dual solution presented in Eqs. \eqref{equation:jet}-\eqref{equation:jet2} and \eqref{equation:Radiojet}-\eqref{equation:Radiojet2} gives a dichotomy in radio properties that is able to explain the distinction between radio-loud and radio-quiet objects. A powerful radio luminosity can be triggered by the two accretion regimens, ADAF and TD. For the former, we need a very massive and spinning BH whose accretion rate $\dot{m}$ needs to be very close to the maximum allowed for an ADAF to occur ($ \sim \dot{m}_c$). For the latter, the radio luminosity does not depend on the accretion rate and it is the BH mass and its spin which play the main role in triggering powerful radio luminosities.

\section{The environment of radio galaxies} \label{sec:Data}

In this section we explore the \texttt{GALFORM} predictions for the typical overdensities around radio galaxies at $z$ = 1.5, 2.2 and 3. We then discuss our results in the context of AGN feedback and compare our theoretical findings with the results of the CARLA survey. 

\subsection{The overdensities around Radio Galaxies} \label{sec:overdensities}
In a hierarchical structure formation scenario,  overdensities around
DM haloes are an increasing function of halo mass
\citep[e.g.,][]{Bardeen1986}. At the same time, more massive galaxies are
expected to live in more massive halos, unless some baryonic process, such as
feedback, is able to prevent stellar growth while dark matter haloes keep accreting  mass \citep{Benson2003}. If AGN feedback prevents the stellar mass build-up of radio galaxies, we expect them to be typically hosted by halos more massive than what the average $M_{\rm Stellar} - M_{\rm Halo}$ relation would predict. Such a difference in the host halo masses of radio galaxies and radio-quiet galaxies with the same stellar mass should thus be reflected in the overdensities around the two populations. 
 
 \begin{figure}
 	\centering
 	\includegraphics[width=68mm]{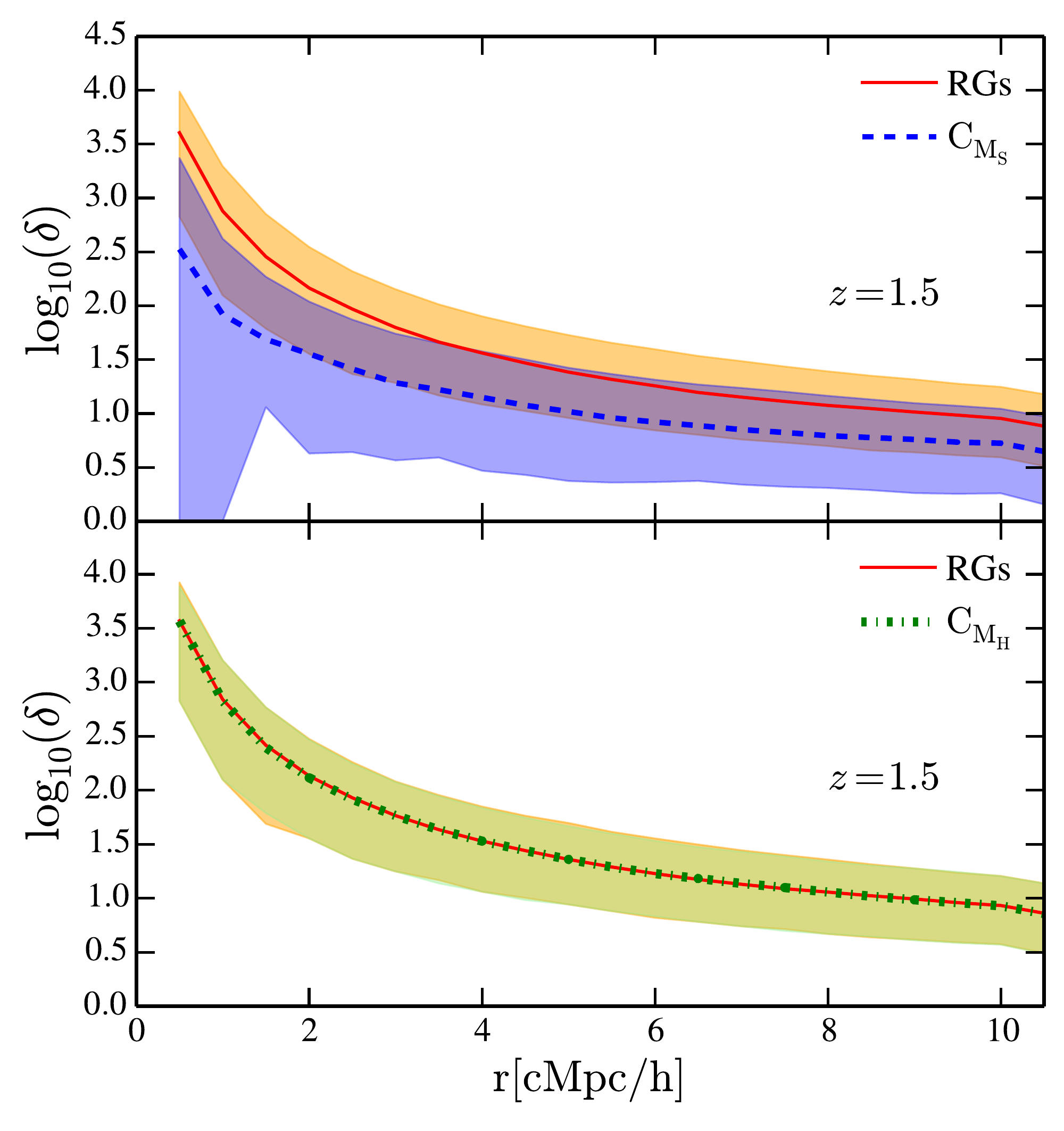}
 	\includegraphics[width=68mm]{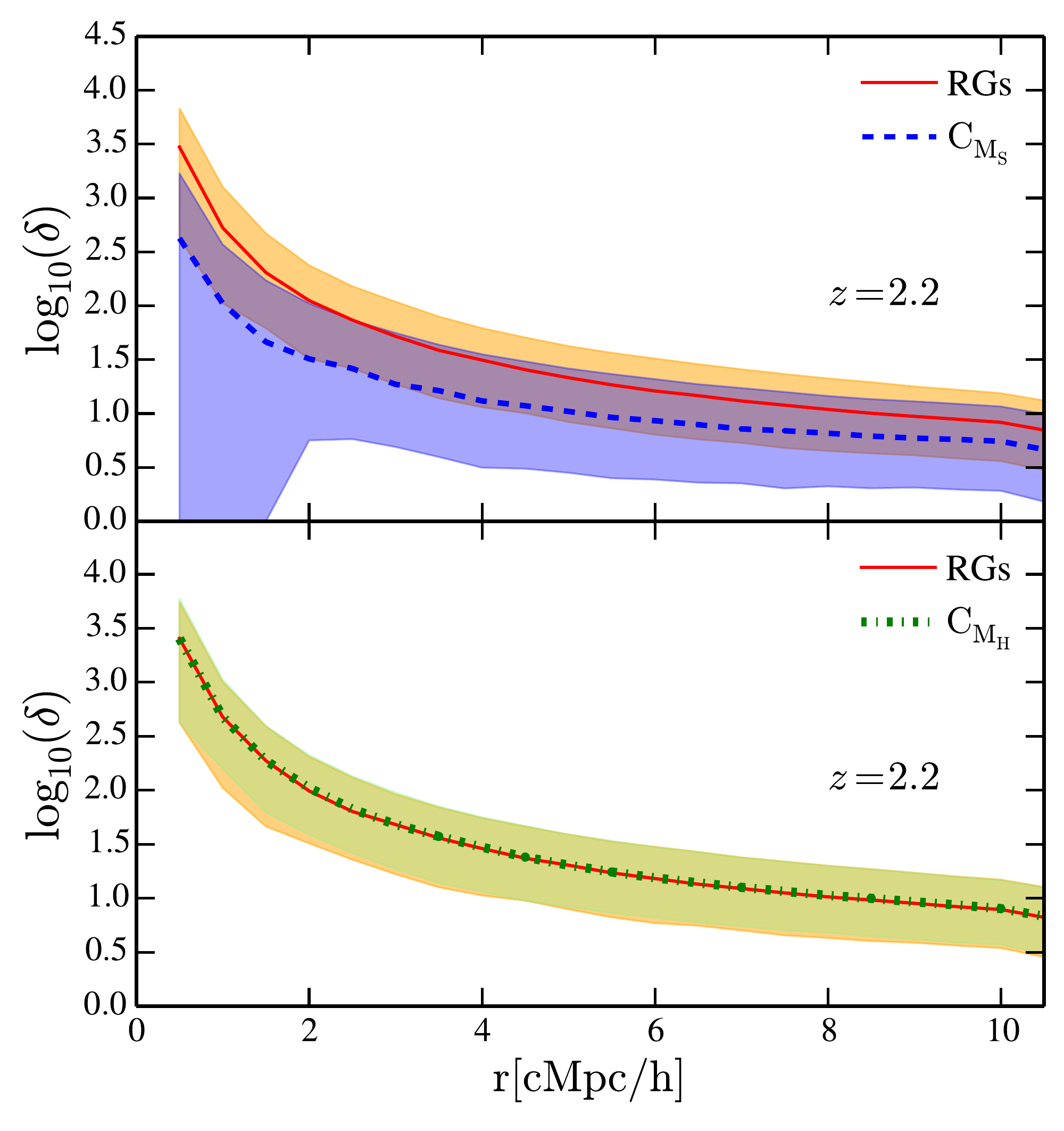}
 	\includegraphics[width=68mm]{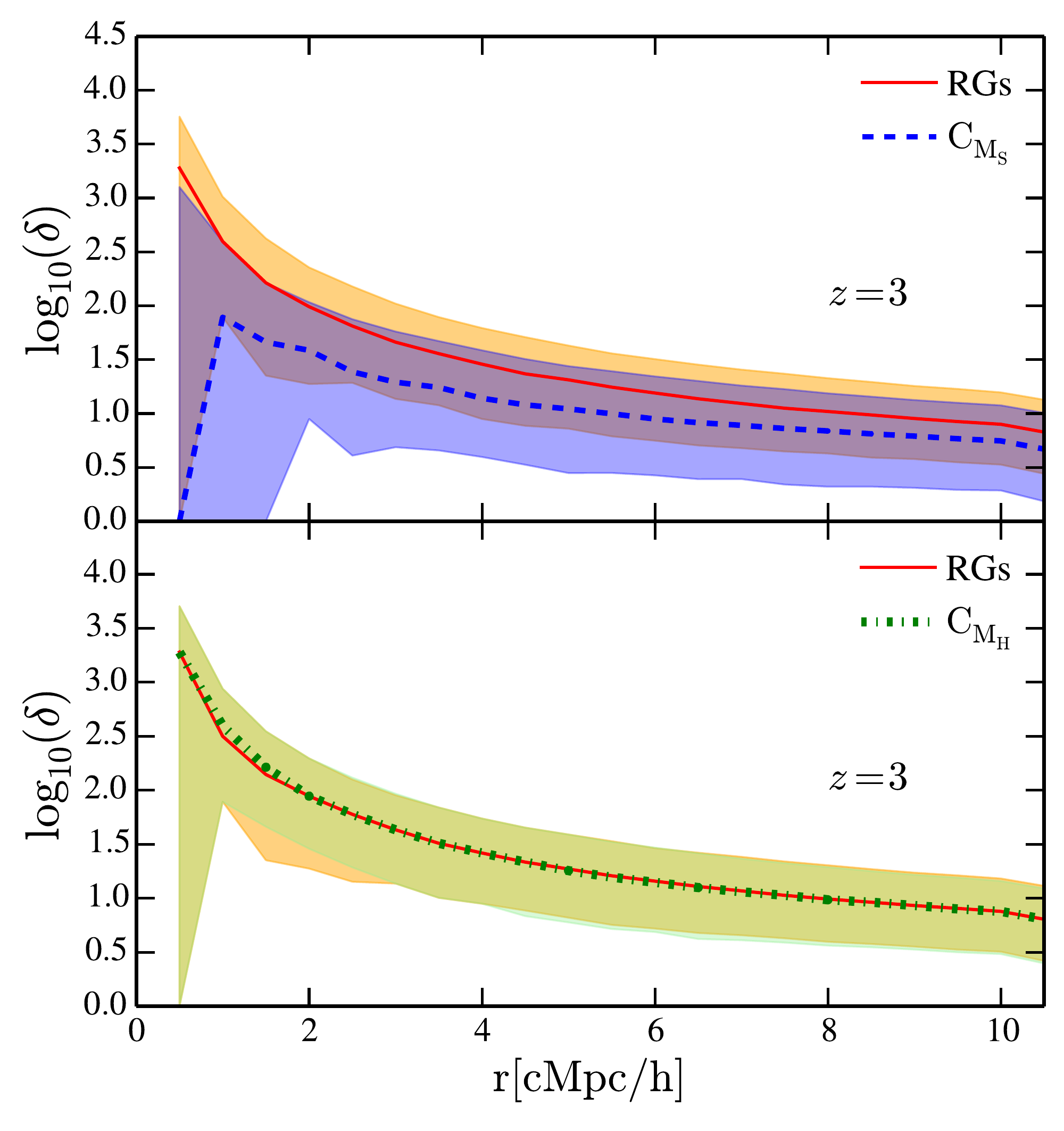}
 	\caption{Galaxy overdensities as a function of distance from radio galaxies (red lines) and the galaxies of the \cm (dashed blue) and \ch (dotted green) control samples. Results are shown for $z = 1.5$ (top), $ z = 2.2 $ (middle) and $z = 3$ (bottom panel). The shading represents the values between the 10 and 90 percentiles (orange for the RGs, blue for the \cm and green for the \ch). }
 	\label{fig:OverdensitiesAGNONN}
 \end{figure}

To test this idea, we select a sample of central\footnote{The term ``central'' refers to galaxies that are located at the center of their host dark matter halo \citep{Springel2005}.} radio galaxies (RGs) from the output of the model, calculate the overdensity of galaxies around these RGs and compare the results against those predicted for two control samples of central galaxies
matched either in stellar mass (\cm) or in host halo mass (\ch). To define the sample of RGs, we choose the 1\% brightest central galaxies in radio luminosity at $\nu$ = 1.4 GHz predicted by \texttt{GALFORM} at each redshift. With this selection, we obtain 1156 objects at $z$ = 1.5, 1806 at $z$ = 2.2 and 2306 at $z$ = 3.0. We explore the model predictions at these three different redshifts, as this is  where the environments of radio galaxies have been characterised observationally \citep{Wylezalek2013,Wylezalek2014, Hatch2014, Cook2015, Cook2016}. Fig.~\ref{fig:Stellardist} shows examples of the spatial distribution of dark matter subhalos and galaxies around two radio galaxies and two galaxies from the \cm sample with comparable 
stellar masses at $z$ = 2.2. RGs are embedded in a dense filamentary web of DM,
where most of the neighbouring galaxies are typically not necessarily satellites
of the central object, but rather close neighbours which may belong to a
different parent halo than the central object. The
green contours in the Figure highlight the projected surface density of
subhalos, showing that RGs are located in denser dark matter regions  than that
of the control sample galaxy.\\

The galaxy overdensity profile $\mathrm{\delta}(r)$ around the radio galaxies and the galaxies of the two control samples is defined as: 
\begin{equation} \label{eq:overdensity}
\delta (r)= \frac{n\,(<r)}{\bar{n}} -1 \; ,
\end{equation}
where $n\,(<r)$ $[\mathrm{Mpc^{-3}} h\rm^3]$ is the number
density of galaxies within a sphere of radius $r$ around the target galaxy and
$\bar{n}$ $[\mathrm{Mpc^{-3}} h\rm^3]$ is the average number density of galaxies
across the simulation box.\footnote{To calculate the number of galaxies around
target objects and the average number density of galaxies across the simulation
box, we included  all galaxies in the simulation with
	$\mathrm{M_{stellar}\geq10^{9}\;M_{\odot}/h}$. We have checked that a
different choice for this lower limit in stellar mass does not affect our results.}

\begin{figure*}
	\centering
	\includegraphics[width=2.16\columnwidth]{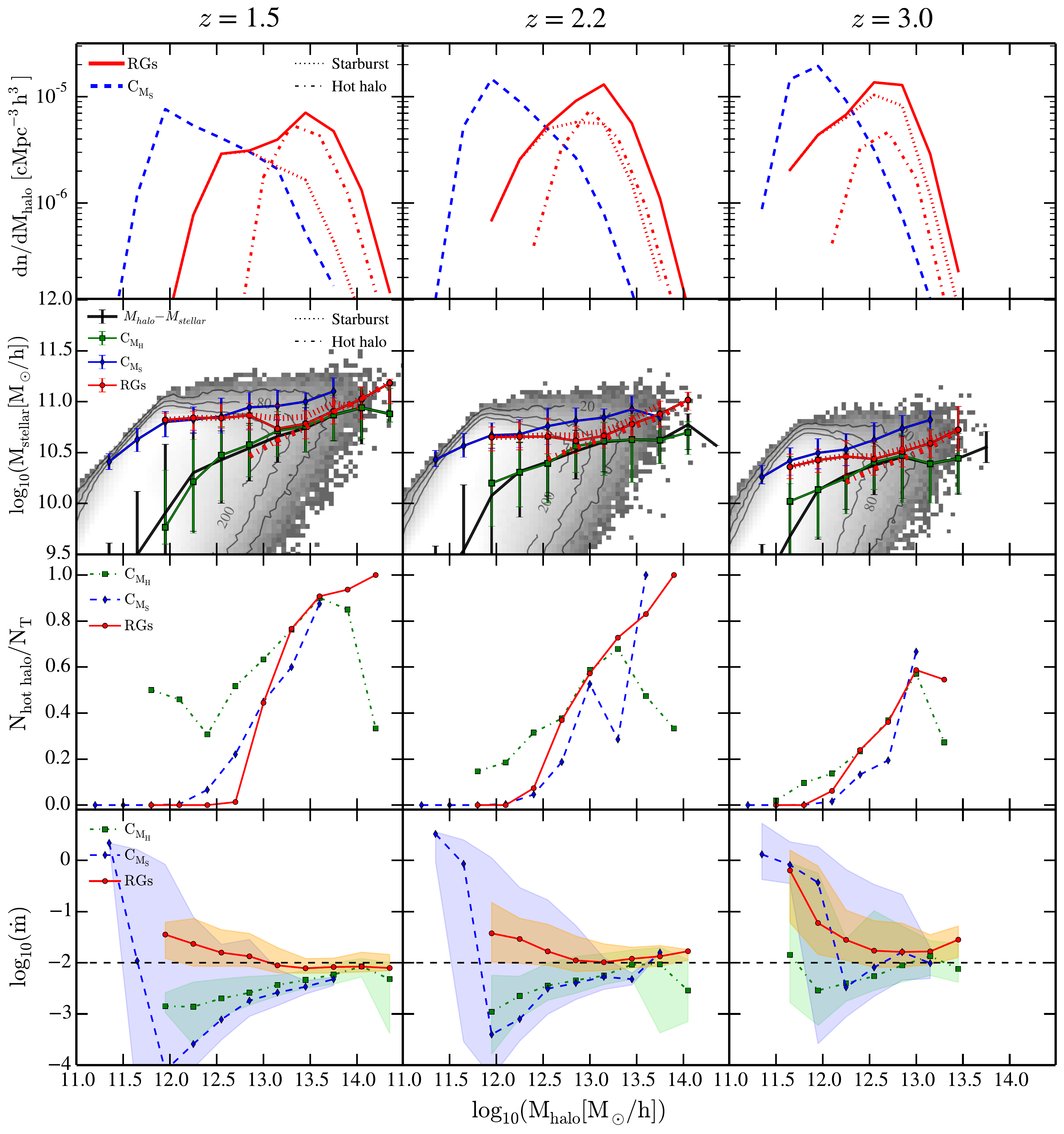}
	\caption{(Top row): Halo mass function of RGs (solid red line) and the \cm galaxies (dashed blue line), at three different redshifts. 
	For the RGs host halos, we show the division between halos whose BHs are accreting gas from the hot-halo (red dash-dotted line) and starburst mode (red dotted line). (Second row): $M_{halo}-M_{stellar}$ relation for central galaxies at $ z  = 1.5$, $2.2$ and $3.0$.The black line corresponds to the median stellar mass per halo mass bins for central galaxies. The red curve shows the same for RGs. Blue diamonds and green squares show this for the \cm and \ch samples, respectively. The bars represent the 16-84 percentile. The red dotted lines represent the relation for RGs whose BHs are accreting cold gas in the starburst mode while the red dashed-dotted lines the same but for the hot halo regime. (Third row): fraction of galaxies per halo bin that are accreting gas from the hot halo atmosphere. RGs are shown in red, \cm in blue and \ch in green. (Bottom row): The median distribution of specific accretion rates ($\dot{m}$) for RGs (red) \cm (green) and \ch (blue) BHs as a function of halo mass.  The dashed line represents the critical AGN accretion rate ($\dot{m}_c$) which separates the two accretion regimes in the model: ADAF and TD. 
	The shaded regions represent the 16-84 percentile range.}
	\label{fig:HMF_Relation_mdot}
\end{figure*}

Fig.~\ref{fig:OverdensitiesAGNONN} shows the predicted galaxy overdensities
around RGs compared to the \cm and \ch samples at the three different redshifts
analyzed. These results correspond to the ideal case in which no projection
effects affect the measured overdensities. Our model predicts that RGs are typically  surrounded by  denser 
environments than galaxies with the same stellar mass distribution. The difference is slightly more pronounced at lower 
redshifts and at small scales ($r \lesssim 2\ {\rm Mpc/}h$), where the median of the distribution associated with radio galaxies is about an
order of magnitude higher than that for the \cm control sample. At larger scales, the overdensity profiles start to converge (and $\delta(r)$ 
will eventually reach zero, by construction, at
scales of several tens of $\mathrm {cMpc}/h$). There are no noticeable
differences between the typical overdensities  around radio galaxies and the galaxies of the \ch sample at any redshift.\\

To identify the mechanism causing the differences in the environment of RGs and the \cm sample, Fig~\ref{fig:HMF_Relation_mdot} upper row shows their halo mass distribution. 
RGs are hosted by more massive halos than the \cm sample, 
with differences that go from $\sim$2 dex  at $z = 1.5$ to $\sim$1.0 dex at $ z = 3$. This is consistent with the results of \cite{Mandelbaum2009}, who found that, at fixed stellar mass, 
radio AGN are found in more massive halos ($\mathrm{1.6\times10^{13} [M_{\odot}}/h]$) than both optical AGN and the bulk of the galaxy population ($\mathrm{\sim 8 \times 10^{11} [M_{\odot}}/h]$).

The differences between RGs and both control samples are also evident in the ${M_{\rm halo}-M_{\rm stellar}}$ plane, shown in the second row of Fig~\ref{fig:HMF_Relation_mdot}. 
The stellar content of central galaxies increases with halo mass, with a scatter of $\sim 1.5$ dex at fixed halo mass. For the most massive haloes ($ \mathrm{M_{halo} \gtrsim 10^{12} [M_{\odot}/}h]$), the relation flattens due to radio mode AGN feedback. 
In the range of halo masses in which the RGs and \cm sample overlap, RGs are hosted by galaxies with lower stellar masses than the \cm galaxies.
On the other hand, the \ch sample matches the median relation of ${M_{\rm halo}-M_{\rm stellar}}$ for the bulk of the galaxy population (by construction), and RGs display higher stellar masses than those galaxies in the \ch sample. 
This implies that  RGs have experienced a different mass assembly history compared to typical galaxies with the same host halo mass. This difference seems to
smear out towards lower redshifts, especially at the massive end, where the main physical process responsible for star formation quenching is AGN feedback. 

In Fig.~\ref{fig:HMF_Relation_mdot} we also display RGs in terms of their main mode of accretion in to the SMBH. 
At all redshifts studied, the RG sample is composed of two different populations, the one who experiences {\it hot-halo} acretion and the one in the {\it starburst} accretion. 
In \texttt{GALFORM}, only the \textit{hot-halo} mode of accretion is linked with the AGN feedback phase \citep[see][]{Bower2006,Lacey2016}.
Cold gas accretion in the \textit{starburst} mode occurs after a merger or a disk instability but does not result in the quenching of star-formation activity, unlike other semi-analytical models like \texttt{GAEA} \citep{Hirschmann2016}. 
RGs experiencing hot halo accretion lie closer to the median relation in the $M_{halo} - M_{stellar}$ plane compared to those experiencing \textit{starburst} acretion. 
The relative abundance of each mode of accretion varies with redshift. Accretion from the \textit{hot-halo} declines towards high redshifts, 
from $\sim 50\%$ at $z=1.5$ to $\sim 23\%$ at $z=3.0$. \textit{Hot-halo} accretion dominates at the massive end ($ M_{halo} \, \mathrm { \gtrsim 10^{12.5} [M_{\odot}} /h]$) of the halo mass function, 
as it is shown in the first and third row of Fig.~\ref{fig:HMF_Relation_mdot}. Despite that not all the galaxies in the \textit{hot-halo} mode are experiencing AGN feedback, in our redshift range these two conditions 
are fulfilled by most haloes with masses above $ M_{halo} \mathrm {\sim 10^{11.8}[M_{\odot}/}h]$ \citep[see Fig. D1 in ][]{Mitchell2016}. Since all our RGs accreting in the \textit{hot-halo} mode 
have $ M_{halo}$ above this threshold regardless of redshift, these are experiencing
AGN feedback. The third row of Fig.~\ref{fig:HMF_Relation_mdot} thus shows, that only a fraction of all RGs are indeed experiencing AGN feedback. 
The remaining RGs are experiencing \textit{starburst} mode accretion. At low halo masses, the fraction of RGs undergoing \textit{hot-halo} accretion is lower 
than that from the \ch sample. At the massive end, \textit{hot-halo} accretion is more common among RGs than in the \ch sample, since the latter also includes galaxies experiencing no accretion at all.

\subsection{Triggering radio galaxies} \label{sec:TrigeringradioLuminosity}

Although there is not a significant difference in the environment of RGs compared to those from the \ch sample, RGs are a sub-sample of galaxies populating massive 
haloes where powerful radio jet emission has been triggered. By comparing both populations, here we explore the mechanisms that allow powerful radio emission in galaxies.

Fig.~\ref{fig:HMF_Relation_mdot} shows that an important fraction of the RGs sample is triggered by mergers and disk instabilities (starburst mode). 
This is consistent with the results of \cite{Chiaberge2015} where 92\% of their radio-loud objects at $z>$1 are associated with recent or ongoing merger events. In \texttt{GALFORM} 
the starburst mode is activated by mergers and disk instabilities that destroy galaxy disks and are responsible for increasing the bulge mass. Thus, most RGs are bulge dominated, 
with a less massive disk component and more massive bulges than the galaxies in the \ch\ sample. Furthermore, we find $\sim\!\!80\%$ of RGs show ongoing 
starburst activity while only $\sim\!\!50\%$ of \ch\ galaxies are actively star forming at $z$ = 1.5, 2.2 and 3.

The specific accretion rate $\dot{m}$ determines the channel of radio power in a galaxy, either via thin disk or ADAF, as discussed in \hyperref[sec:GALFORM]{Section~\ref{sec:GALFORM}}.
The bottom panels of Fig~\ref{fig:HMF_Relation_mdot} show the median $\dot{m}$ per bin of halo mass for RGs, \cm and \ch. RGs hosted in the most massive 
halos are characterized by the ADAF channel  with $\dot{m}$ values distributed around $\dot{m}_c$ (i.e. the maximum radio jet luminosity 
allowed in the ADAF channel). On the other hand, RGs hosted by less massive halos, fuelled by starburst accretion, have $\dot{m}$ values bigger
than $\dot{m}_c$ so that their BH accretion is characterized by the TD channel. Regarding the control samples, \ch is always 
characterized by an ADAF with lower Eddington rates than RGs. The \cm median accretion rates are overall associated with the ADAF channel.

In summary, RGs are triggered by both TD and ADAF channels which are fuelled by hot and cold gas from the hot halo atmosphere and mergers/disk instabilities, respectively. 
The TD channel can trigger a powerful radio-loud jet when the BH in the host RG is very massive ($\mathrm{M_{BH}\gtrsim 10^{8.5}[M_{\odot}/}h]$) and the BH spin is around $a\sim0.5$. 
For the ADAF channel, in addition to these constraints, the accretion rate needs to be close to the maximum allowed for an ADAF to occur. Furthermore, RGs triggered by 
cold gas accretion (\textit{starburst} mode) are more abundant at $z \gtrsim 2.2$ while at lower redshifts, hot gas accretion becomes the main mode of RGs triggering.

\subsection{The effects of AGN feedback on radio galaxies} \label{subsec:AGNEffect}
As we already discussed, since radio power is linked to AGN activity, it is
likely that feedback from accreting black holes is the main physical
mechanism  responsible for quenching star formation in RGs, thus causing these galaxies to lie in the $M_{halo}-M_{stellar}$ plane closer to the galaxies that have already experienced significant quenching (the massive end).  
To verify that  AGN feedback is responsible for the quenching of the most massive RGs, we run a variant of \texttt{GALFORM} in which AGN feedback is ``switched off'', and study the environment of the galaxies that were classified as RGs in the
original run. The counterparts of RGs in the new run (hereafter \RGo) are found by matching the
position and halo masses of the galaxies in the new run with the ones of RGs in
the original run.

\begin{figure}
	\centering
	\includegraphics[width=1\columnwidth]{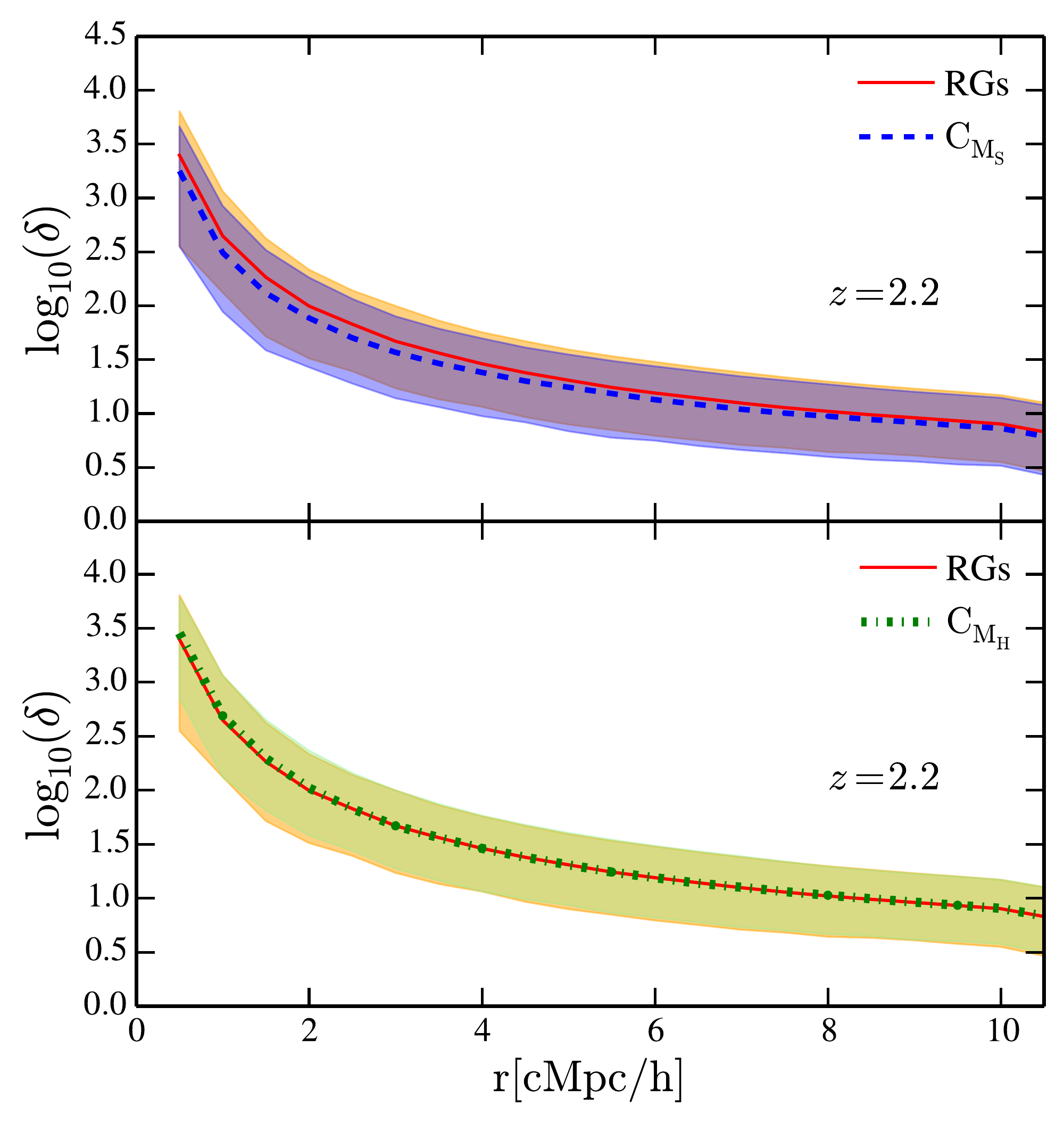}
	\caption{Overdensity around central galaxies when AGN feedback is switched off. The shading represents the values between the 10 and 90 percentiles (orange for RGs, blue for \cm and green for \ch). The two plots show an overdensity comparison around RGs (red), \cm (dashed blue in top panel) and  \ch (dotted green in the panel below) at $ z = 2.2 $.}
	\label{fig:OverdensityAGNoff}
\end{figure}

We then generate also two new control galaxy samples, matching the stellar masses and the halo masses of the \RGo. We refer to the new control samples with \cmo and \cho, respectively. We then calculate the overdensities around the new three population showing in Fig.~\ref{fig:OverdensityAGNoff} the results for $z=2.2$. In contrast with the  results of the original run, now the overdensity  profiles around the RGs counterparts and the \cmo sample (upper panel) are almost identical, indicating that now there are no important differences in the halo masses hosting \RGo and the control sample matched in stellar mass. Consistently, the overdensities around the \RGo sample and  the \cho sample (lower panel) are  found to be undistinguishable, as in the original model.

This simple test run confirms the idea that the AGN feedback shapes the RGs overdensities.

\subsection{Comparison with observations} \label{sec:CARLA}
The previous section showed that the environment of radio galaxies can provide
information about the effect of AGN feedback on the stellar content of these
objects. Here, we compare our model predictions of the overdensities around radio
galaxies with recent  observational results from the CARLA survey
\citep{Wylezalek2013,Wylezalek2014,Cook2015, Cook2016}.\\

The CARLA survey is a warm \textit{Spitzer} program designed to study the
environment of nearly $400$ radio-loud sources, of which $187$ are radio-loud
quasars and $200$ are radio galaxies. Targets were selected over the redshift
range $1.3 <z< 3.2$. The CARLA AGN sample is composed of powerful RLAGN whose
luminosity at 500 MHz is above $\mathrm{10^{27.5} \; W\;Hz^{-1}}$. In order to
compare environmental properties, the CARLA team selected  a control sample from UKIDSS Ultra Deep Survey (UDS) \footnote{UKIRT Infrared Deep Sky Survey (UKIDSS, \citealt{Lawrence2007}) Ultra Deep	Survey (UDS) is a near-infrared survey covering 0.77 $\rm deg^2$ in the $J$, $H$ and $K$ bands \citep[see][]{Almaini2017}.} 
composed of radio quiet galaxies \footnote{galaxies with radio
luminosities at least two orders of magnitude lower than radio luminosities of CARLA RLAGNs.} with the same stellar mass and redshift distribution. To study the environment of the radio loud and quiet sources, they relied on  IRAC colour-selected
galaxies using IRAC channel 1 (IRAC1) and 2 (IRAC1) with effective wavelengths
of 3.55 $\rm \mu m$ and 4.49 $\rm \mu m$ respectively. Specifically, {\it Spitzer}-selected
sources were defined as either sources brighter than the IRAC2 95\% completeness
limit, above an IRAC1 flux of 2.5$\mu$Jy (3.5 $\sigma$ detection limit) and with
a colour of [3.6]-[4.5] $>$ -0.1 or as sources detected above the IRAC2
95\% completeness limit, an IRAC1 flux $<$ 2.8 $\mu$Jy and a colour $>$-0.1 at
the 3.5$\sigma$ detection limit of the IRAC1 observation
\citep{Wylezalek2014,Hatch2014}. This means that all IRAC-selected sources are
95\% complete in the IRAC2 band down to [4.5] = 22.9, but are not necessarily
detected in IRAC1. The cuts were performed in order to get an homogeneous sample
of field galaxies between $1.3 <z< 3.2$ with a 10\% -20\% contamination level by
low redshift interlopers \citep{Muzzin2013b}.\\
\begin{table}
	\Large
	\centering
	\def\arraystretch{1.35}
	\resizebox{6.5cm}{!}{
		\begin{tabular}{ccccc} 
			$\boldsymbol{z}$ & $\boldsymbol{\mathrm{N_{RG}}}$ & $\boldsymbol{\mathrm{L_ {\nu_{500MHz}}^{min}}}$\textbf{[W/Hz]}& $\boldsymbol{\mathrm{L_ {\nu_{500MHz}}^{max}}}$ \textbf{[W/Hz]} & $\boldsymbol{\mathrm{N_{C_{M_{S}}}}}$  \\ \hline \hline
			\textbf{1.5} & 1969  & 9.30 $\times$ $\mathrm{10^{25}}$ &  2.45 $\times$ $\mathrm{10^{27}}$  & 1969  \\ 
			\textbf{2.2} & 2077  & 8.32 $\times$ $\mathrm{10^{25}}$&  1.57 $\times$ $\mathrm{10^{27}}$  & 2077 \\
			\textbf{3}  & 1897   & 4.95 $\times$ $\mathrm{10^{25}}$&  2.45 $\times$ $\mathrm{10^{27}}$    & 1897  \\ \hline \hline

		\end{tabular}
	}
	\caption{Number of galaxies in RG and \cm samples and the radio luminosity range that RGs display at $\nu = 500\,\rm MHz$ at $z = 1.5$ at $z = 2.2$ and $z = 3$.}
	\label{table:newRGs}
\end{table}

To compare with the CARLA results, we selected the $0.1\%$
brightest radio sources at $500$ MHz in the \texttt{GALFORM}
outputs. We choose this percentage in order to have a large sample of RGs ($\rm
\sim 10$ times bigger than the number used in the CARLA survey) and, at the same
time, selecting only the most powerful radio galaxies in the model. As for the
control galaxies, we built a galaxy sample with the same stellar mass
distribution as the selected radio galaxies. The number of RGs and galaxies in
the control sample at each redshift ($z=1.5$, $2.2$, $3$) is presented in
\hyperref[table:newRGs]{Table~\ref{table:newRGs}}. In order to reproduce the
same redshift distribution for the radio sources as the CARLA sample (shown in
\citealt{Hatch2014}), we randomly selected the same number
of objects in each redshift bin, using our RGs and control sample at redshift $z=1.5$ to cover the observed range $[1.3 - 1.8]$, the $z = 2.2$ sample for the range $[1.8 - 2.5]$
and the $ z = 3$ sample for $[2.5 - 3.1]$.

The sample of field galaxies used to estimate overdensities is selected with the
same constraints as are applied in the CARLA survey. To account for the projection
effects in observations, we join and stack our three redshift boxes along the z-axis
computing the projected overdensities in the same redshifts bins analysed by
CARLA. In order to remove a large fraction of interlopers, \citet{Hatch2014}
excluded sources with [4.5]$<$19.1 mag following the results from
\cite{Wylezalek2014}. To make a fair comparison between our predictions and
their observed results, on top of  applying the color selections explained
above, we impose  this last magnitude cut to select field galaxies.

\begin{figure}
	\centering
	\includegraphics[width=1\columnwidth]{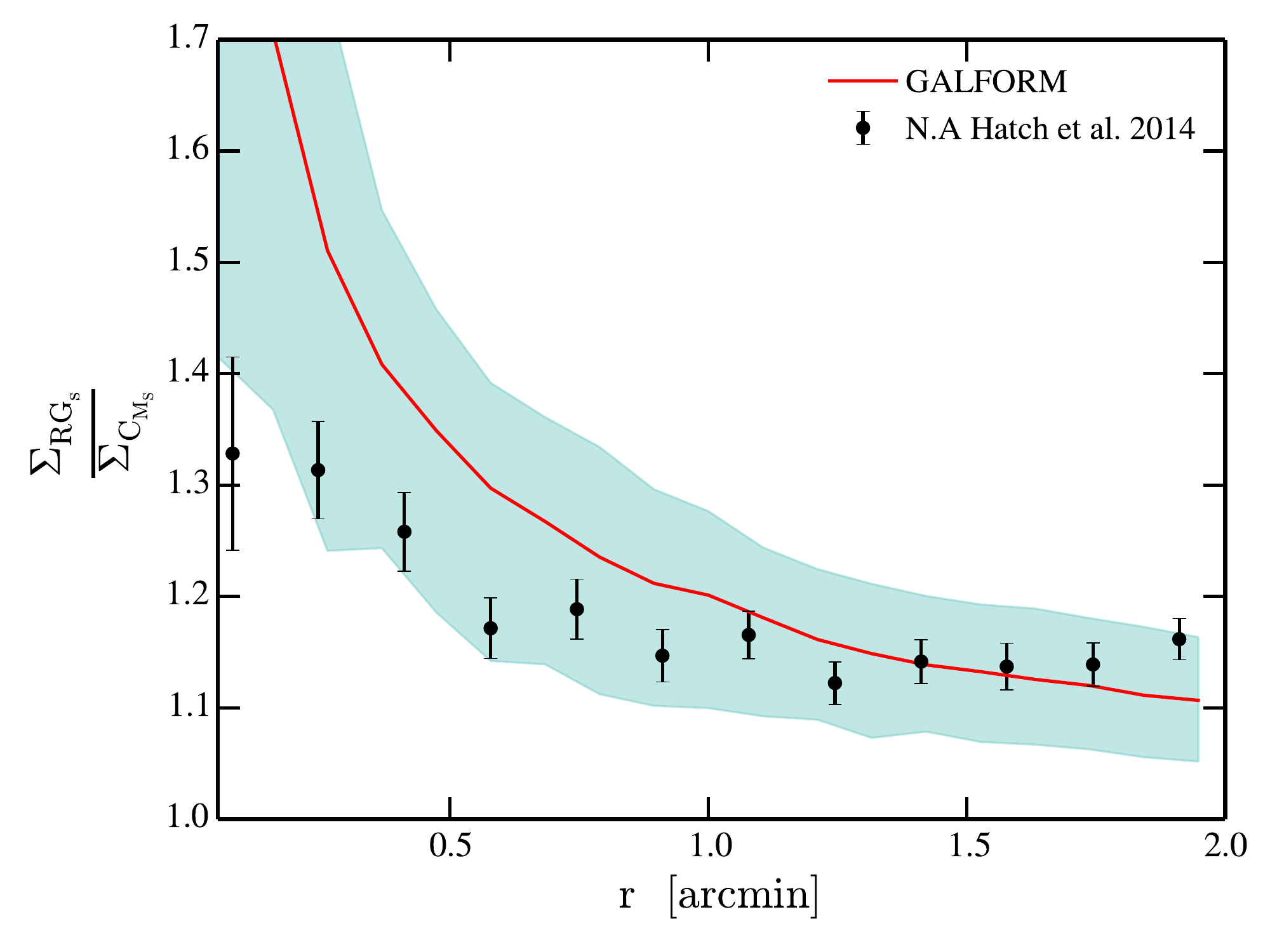}
	\caption{Ratio of the surface density of objects around radio galaxies
and around control sample. The red curve indicates the median of the
\texttt{GALFORM} mocks predictions, with the 10-90 percentiles shown with the
blue area. Black dots: the ratio of radio galaxies and control sample projected
overdensity found in \protect\cite{Hatch2014}. To guide the reader, for the
cosmology assumed in the simulation, $1$ [arcmin] corresponds to $\sim 0.5$
proper Mpc  at $z = 2.2$ (or  1.63 comoving Mpc).}
	\label{Fig:cociente}
\end{figure}

The large volume of the P-Millennium simulation allows us to create  $33$
different mocks that mimic the observational selection of the CARLA survey,
which we  use to study the impact of cosmic variance. 
In Fig~\ref{Fig:cociente} we show the ratio of the projected overdensities
around radio galaxies and the control sample in the simulation, and compare them
with the CARLA survey data. The ratio of projected overdensities measured in the
CARLA sample (black dots) is consistent (within the 10-90 percentile range, blue
shading in the figure) with our model predictions for $\theta \gtrsim 0.4 \,
[{\rm arcmin}]$ ($\sim$200 proper kpc at $z$ = 2.2).  
The slight tension at smaller angular distances can be explained as a result of
observational effects like cosmic variance or due to the fact that
\texttt{GALFORM} predicts that RGs are hosted by very massive halos
\citep{Fanidakis2013,Orsi2015}. However, less massive halos could fuel powerful
radio jets by a variety of mechanisms such as a magnetohydrodynamic acceleration near
to the BH or a transition between ADAF and TD accretion flows
\citep{BlandfordANDZnajek1977,WilsonANDColbert1995,Meier2002,Sikora2007,Beckwith2008}.
Nevertheless, the remarkable agreement shown in Fig.~\ref{Fig:cociente} suggests that our physical interpretation of the environments of radio galaxies is a clear signature of the impact of AGN feedback at these high redshifts.

One of the limitations of the CARLA results is the uncertain redshift of the
sources used to trace the environments. Ideally, a multi-object spectrograph
could be used to systematically identify galaxies at the same redshift as the
central object. However, this has only been performed for a small numbers of objects at most \citep[see
e.g.][]{Overzier2001,Venemans2004,Venemans2005,Geach2007,Kuiper2011}. The future
survey WEAVE-LOFAR \citep{Smith2016}, which combines radio selection with the
new WEAVE spectrograph, will provide a large sample to study the influence of
powerful AGN on their surroundings. Alternatively, narrow-band photometry
targeting emission-line objects can detect galaxies in a narrow redshift window
that matches the objects of interest. The J-PAS photometric survey
\citep{Benitez2014}, for example, is expected to map $\sim 8000 {\rm \, deg^2}$ of the northern
sky with multiple narrow and broad-band filters with a redshift accuracy of
$\sigma_z \approx 10^{-3} (1+z)$.\\

\section{The quenching of STAR-FORMATION AROUND RADIO GALAXIES HALOS} \label{sec:Quenching around radio galaxies}
\begin{figure*}
	\centering
	\includegraphics[width=1.\textwidth]{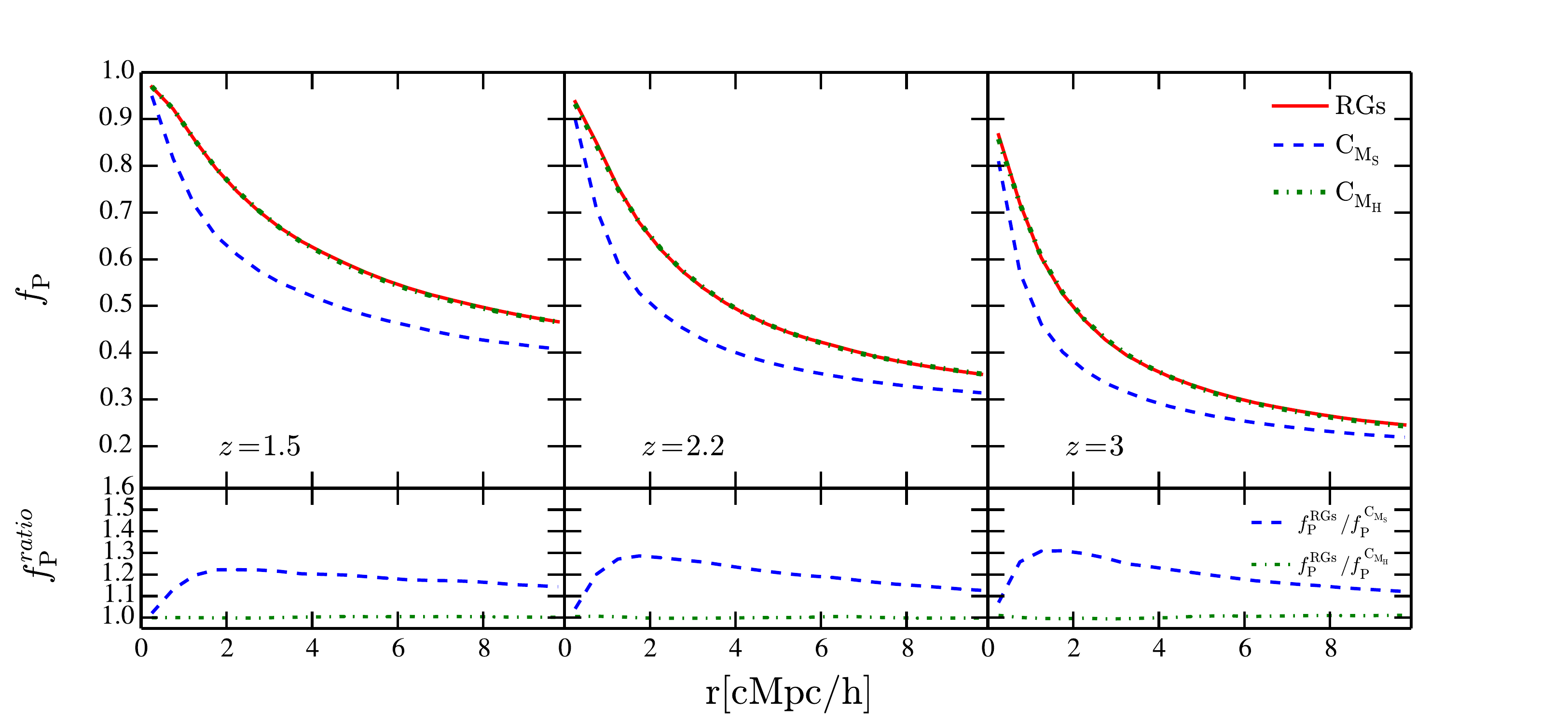}
	\caption{Top panels show the fraction of passive ($f_{\rm P}$) galaxies around RG and control samples (dashed blue line for \cm and dash dotted green line for \ch). The left panel is for $z = 1.5$, the middle one is for $z = 2.2$ and the right one for $z = 3 $. The bottom panels show the ratio of fraction of passive galaxies between RGs and the two control samples: \cm ($ f_{\rm P}^{ratio} \mathrm{\;=\;} f_{\rm P}^{\rm RGs}$ / $f_{\rm P}^{\rm C_{M_S}}$) and \ch ($f_{\rm P}^{ratio} \mathrm{\;=\;} f_{\rm P}^{\rm RGs}$ / $f_{\rm P}^{\rm C_{M_H}}$).}
	\label{fig:FractionPassiveStarburst}
\end{figure*}

The previous section shows that the properties of RG environments can be
explained if these galaxies are hosted by DM haloes about an order of magnitude more massive than their radio-quiet counterparts. In such dense environments, we also expect to find differences in the properties of the galaxies around radio galaxies and the radio-quiet control sample. Environmental mechanisms, such as ram-pressure stripping of  gas, have a greater impact in massive haloes \citep{Dressler1986,Goto2003,Heinz2003,Fujita2004,Roediger2009}.

In order to explore this further, we look at the fraction $f_{\rm P}$ of passive galaxies around RGs and in the control samples. We define ``passive'' as those galaxies whose specific star formation rate (sSFR) is lower than $10^{-10}\ {\rm yr^{-1}}$. In Fig~\ref{fig:FractionPassiveStarburst} we show $f_{\rm P}$  computed for both RGs and  the control samples as a function of distance $r$ to the central object. RGs, the \cm and \ch samples converge to the same $f_{\rm P}$ at $r<0.25 \ {\rm Mpc/}h$ meaning that, at these small distances, galaxies are hosted inside the main halo and have therefore undergone the same physical processes. At $z= 3.0$ the median virial radius of RGs halos is 560 [kpc/$h$], at $z= 2.2$ is 775 [kpc/$h$] and at $z = 1.5$ is 1000 [kpc/$h$] while the median virial radius of the \cm sample is roughly constant in the three redshift bins with a value of 350 [kpc/$h$].At $r>0.25 \ {\rm Mpc/}h$, $f_{\rm P}$ for the \cm sample shows a significant deviation from the behaviour of RGs and \ch fractions, which are identical at every redshift. We stress that the quenching around RGs and the \ch sample is only due to their typical halo mass and not due to the AGN feedback which only affects the central galaxy.

The bottom panels in Fig~\ref{fig:FractionPassiveStarburst} show the radial profiles of the ratio $f_{\rm P}^{ratio}$ between the values for RGs and \ch and RGs and \cm $f_{\rm P}$. While the former is flat at any $r$, the latter increases up to a peak value. At high $r$, the $\Delta f_{\rm P}$ between \cm and RGs reaches $1$, showing that the fraction of passive galaxies in RGs and the \cm sample converges eventually to the average fraction of passive galaxies in the box. These panels also show that the position of the $f_P ^{RGs} \mathrm{/} f_P ^{C_{M_s}}$ peak depends slightly on redshift. This result is related to the increase of the virial radius of halos hosting RGs with time.\\

We further study the properties of the galaxies surrounding RGs by looking at
their infrared luminosity function  (IR LF), which we can compare with the data from the
CARLA survey. Specifically, we compute the total observed IR LF by selecting galaxies inside a radius of 1 arcmin centred on the RGs that meet the selection
criteria set out for CARLA objects in
\hyperref[sec:CARLA]{section~\ref{sec:CARLA}}. We then measure the IR LF in the
three redshift intervals and we compare them with the measurements of
\citet{Wylezalek2014}. To  account for projection effects in the observations,
we stack our three redshift boxes spanning $1.5<z<3$ along the z-axis. In order to mimic the background subtraction of \cite{Wylezalek2014}, we place 500 random and non overlapping apertures with 1 arcmin of radius onto our projected field to estimate the typical field density of IRAC-selected sources. Then, we compute the average blank field LF and we subtract it from the luminosity function of each RGs. To estimate cosmic variance, we randomly choose 30 RGs from \hyperref[table:newRGs]{Table~\ref{table:newRGs}} for each redshift bin to mimic the CARLA sample of RGs.\\

\begin{figure*}
	\centering
	\includegraphics[width=170mm]{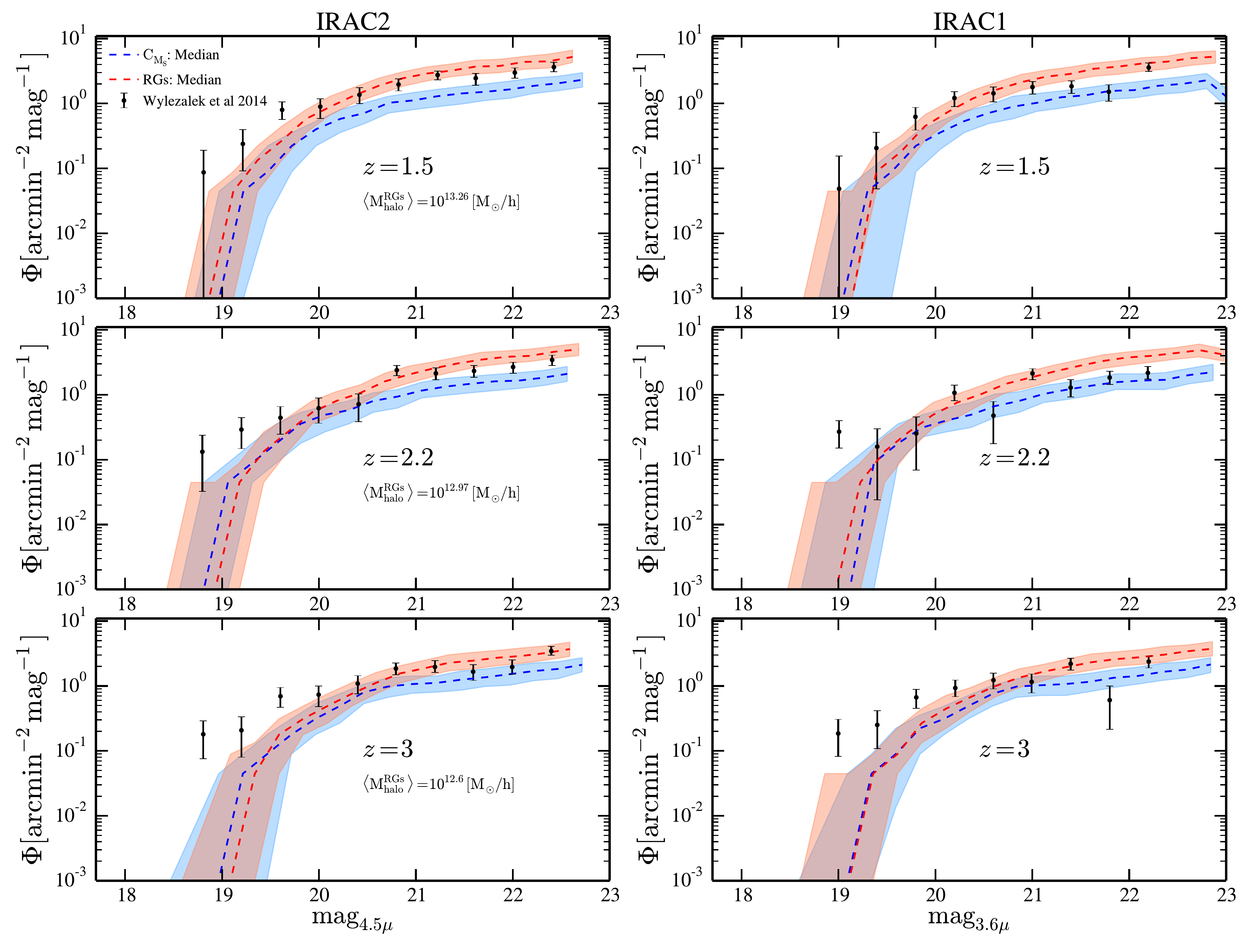}
	\caption{Dashed lines are the LFs in IRAC bands predicted by \texttt{GALFORM} for RGs (red) and \cm (blue) environments, the red and blue shaded regions represent the 10-90 percentiles of the mocks for RGs and \cm respectively. The black points show the observational LF of CARLA sample analysed by \protect\cite{Wylezalek2014}. The right column corresponds to the apparent magnitude of sources in the IRAC band 1 with an effective wavelength of 3.55 $\rm \mu m$ and left column the same but in the IRAC band 2 with an effective wavelength of 4.49 $\rm \mu m$.
	The two upper panels show the results in the snapshot redshift $z = 1.5$ and CARLA sample data in the redshift interval 1.5 $< z <$ 1.9. The middle panels show are the results in the snapshot redshift $z = 2.2$ and the CARLA sample data in the redshift interval 2.1 $< z <$ 2.3. And, last bottom panels the same for the $z = 3$ snapshot and the CARLA sample interval 2.6 $< z <$ 3.1.}
	\label{fig:LF_CARLA}
\end{figure*}

The results for the predicted IR LF of galaxies around RGs are shown in
Fig~\ref{fig:LF_CARLA}, together with the measurements from the CARLA survey. The right column corresponds to the luminosity function around RGs and \cm in the IRAC 1 band with an effective wavelength of 3.55 $\mu m$ and the left column the same but in the IRAC 2 band with an effective wavelength of 4.49 $\mu m$. Our results are in good agreement with the observations throughout the redshift range $1.5<z<3$, except for the brightest bins at the highest $z$. However, there is a remarkable agreement between the faint end of the observed and predicted LFs suggesting that our predicted RGs environments are consistent with
the observed ones.

We note that for all redshifts the faint end of the RGs LFs is higher than that for \cm. The fraction of passive galaxies is higher in the RGs environments than for \cm ones. Hence, the  \textit{Spitzer} IR LF of objects around the \cm sample is significantly below the LF of RGs environment, suggesting that the abundance of passive galaxies around these environments is consistent with the observational measurements.

\section{Summary and CONCLUSIONS} \label{sec:Summary}
Recent observations of radio galaxies (RGs) have suggested that their environments are denser than those of their radio-quiet counterparts \citep{Hatch2014}. To understand this phenomenon from a theoretical perspective, we use the semi-analytical model of galaxy formation \texttt{GALFORM} \citep{Lacey2016}. This model features a detailed modelling of the co-evolution of galaxies and their central SMBH, including its growth through different accretion channels, spin evolution and the regulation of star formation through AGN feedback. 

We explore the model predictions at redshift $1.5$, $2.2$ and $3$ in which most of the observational work about RGs environments has been carried out. In order to analyse the overdensities around radio galaxies we construct two types of control samples, one with the same stellar mass distribution (\cm) and the other one with the same halo mass of as the RGs (\ch). Our model predictions are consistent with RGs being in denser environments than galaxies from the \cm sample, in similar environments to the \ch sample. The latter suggests that the ovserdensities around RGs are determined solely by their host halo masses. In fact, RGs are hosted by massive haloes ($\mathrm{10^{11.75}} < M_{halo} < \mathrm{10^{14} \; [M_{\odot}}/h]$) which are on average $\sim$1.5 dex more massive than those from the \cm sample.\\

Given that RGs are preferentially hosted by very massive halos, in which AGN feedback is the most common mechanism to quench star formation \citep{Bower2006,Croton2006}, we expect that the RGs stellar mass build-up process was slowed down due to the black hole feedback. This would make them be hosted by halos with less stellar mass with respect to the predicted by the $M_{stellar} - M_{halo}$ relation. In order to test this idea as the main driver of the differences in overdensities around RGs and \cm, we ran a variant of our model in which  AGN feedback is switched off. We find that the halo mass distributions of RGs and the \cm sample are comparable and they reside in similar environments. These results corroborate the idea that radio galaxies have less stellar mass due to AGN feedback effectively preventing star formation.

Interestingly, we found that while the \ch sample follows the median relation of ${M_{\rm halo}-M_{\rm stellar}}$ for the bulk of the galaxy population, RGs lie systematically above this relation, implying that  RGs have experienced a different mass assembly history compared to typical galaxies with the same host halo mass. This difference seems to
smear out towards lower redshifts, especially at the massive end, where the main physical process responsible for star formation quenching is AGN feedback. To explore the role of this process in shaping the stellar content of radio galaxies, we split them into the ones currently in the \textit{hot halo} accretion mode and the ones in the \textit{starburst} mode. In the \texttt{GALFORM} model only the \textit{hot halo} mode is linked with the AGN feedback \citep[see][]{Lacey2016}. We found that the RG sample is composed of galaxies in these two accretion modes whose relative proportion varies with redshift. Around $\sim$ 23.7\% $z = 3.0$, $\sim$ 43\% at $z = 2.2$ and $\sim$ 53.6\% at $z = 1.5$ of RGs black holes show accretion from the surrounding hot gas atmosphere (i.e, \textit{hot halo} mode) as the main channel of growth being the principal mode at higher halo masses ($ M_{halo} \, \mathrm {\gtrsim 10^{12} [M_{\odot}} /h]$). Only these galaxies, among all RGs, are experiencing feedback from the black hole. On the other hand, the other sample of RG black holes is growing due to cold gas accretion in the \textit{starburst} mode after a merger or a disk instability but is not linked with an AGN feedback phase. Therefore, only a fraction of all RGs are indeed experiencing AGN feedback: at low halo masses, the fraction of RGs experiencing \textit{hot-halo} accretion is low while in the massive end the \textit{hot-halo} accretion is more common among RGs. 
In addition, the two modes in RGs are preferentially associated with different accretion flows geometries: ADAF is linked to the \textit{hot halo} mode whereas TD accretion flows are dominant in the \textit{starburst} mode. Both cases, at any redshift, exhibit $\dot{m}$ values close to the critical threshold $\sim 0.01$ in Eddington units.

The versatility of the \texttt{GALFORM} model allows us to compare the predictions against observational measurements of the environments of RGs. We build mock catalogues of the CARLA RG sample \citep{Wylezalek2013,Hatch2014}. We find remarkable agreement between model predictions and observations when comparing the density of objects around the RGs and the control sample. This supports the physical picture explored here in which the comparison of environments of RGs and their radio-quiet counterparts reveals the effect of AGN feedback at these high redshifts.

Since RGs are hosted by more massive haloes than their radio-quiet counterparts, we expect that the galaxies in their environments experience different tranformation mechanisms. This is reflected in the relative fraction of passive galaxies around RGs and around galaxies in the \cm sample. Galaxies surrounding the \ch sample have the same fraction of passive as those around RGs. This means that the halo mass distribution is the main property determining the fraction of passive galaxies in the environment of a central object. To validate our model predictions, we compare the observed infrared luminosity function in the \textit{Spitzer} bands (IRAC1 and IRAC2) of RG environments at different redshifts with the observational measurements of the \textit{Spitzer} IR LF in the CARLA survey shown in \cite{Wylezalek2014}. The model shows remarkable agreement with the observational data throughout  the redshift range $1.5<z<3$, except at the brightest bins at the highest redshifts.\\ 

Current data samples of environments of high redshift RGs typically suffer from sample and cosmic variance due to their small size. However, the agreement with our model predictions is encouraging and suggests that future more ambitious observational campaigns could be designed to put constraints on the strength of AGN feedback at high redshifts. For instance, the J-PAS survey is expected to map $\sim8000 {\rm \, deg^2}$ of the northern sky with multiple narrow and broad-band filters with a redshift accuracy of $\sigma_z \approx 10^{-3} (1+z)$ \citep{Benitez2014}. Such a data sample, cross-matched with a high-redshift RG catalogue, would allow us to characterise the environments of these objects with unprecedented accuracy. Likewise, forthcoming multi-object spectroscopic surveys, such as the WEAVE-LOFAR survey \citep{Smith2016} are expected to increase the number of known RGs at high redshifts and characterise their environments. By comparing the results from these large data samples to galaxy formation model predictions such as the ones presented here, we expect to be able to put tight constraints on the physical mechanisms regulating galaxy formation and evolution at high redshifts.
\section*{Acknowledgements}
We acknowledge encouraging discussions with Dominika Wylezalek that helped shaping the ideas presented in this paper. We also acknowledge support from project AYA2015-66211-C2-2 of the Spanish Ministerio de Economia, Industria y Competitividad and also STFC Consolidated Grants ST/L00075X/1 and ST/P000451/1 at Durham University. This work used the DiRAC Data Centric system at Durham University, operated by the Institute for Computational Cosmology on behalf of the STFC DiRAC HPC Facility (www.dirac.ac.uk). This equipment was funded by BIS National E-infrastructure capital grant ST/K00042X/1, STFC capital grants ST/H008519/1 and ST/K00087X/1, STFC  DiRAC  Operations  grant  ST/K003267/1  and Durham University. DiRAC is part of the National E-Infrastructure.

\bibliographystyle{mnras}
\bibliography{references}

\bsp	
\label{lastpage}

\end{document}